\documentclass[conference]{IEEEtran}
\IEEEoverridecommandlockouts

\usepackage{cite}
\usepackage{amsmath,amssymb,amsfonts}
\usepackage{algorithm}
\usepackage{algpseudocode}
\usepackage{graphicx}
\usepackage{textcomp}
\usepackage{multirow}
\usepackage{booktabs}  
\usepackage{xcolor}
\def\BibTeX{{\rm B\kern-.05em{\sc i\kern-.025em b}\kern-.08em
    T\kern-.1667em\lower.7ex\hbox{E}\kern-.125emX}}
\begin{document}

\title{Learning to Discover: A Generalized Framework for Raga Identification without Forgetting
\\ 
}

\author{
\IEEEauthorblockA{
Parampreet Singh$^{\dagger}$, Somya Kumar$^{\dagger}$, Chaitanya Shailendra Nitawe$^{\dagger}$, Vipul Arora$^{\dagger \ddagger}$\\
$^{\dagger}$Indian Institute of Technology Kanpur, $^{\ddagger}$Katholieke Universiteit Leuven\\
}
}

\maketitle

\begin{abstract}
Raga identification in Indian Art Music (IAM) remains challenging due to the presence of numerous rarely performed
Ragas that are not represented in available training datasets. Traditional classification models struggle in this setting, as they assume a closed set of known categories and therefore fail to recognise or meaningfully group previously unseen Ragas.
Recent works have tried categorizing unseen Ragas, but they run into a problem of catastrophic forgetting, where the knowledge of previously seen Ragas is diminished.
To address this problem,
We adopt a unified learning framework that leverages both labeled and unlabeled audio, enabling the model to discover coherent categories corresponding to the unseen Ragas, while retaining the knowledge of previously known ones. We test our model on benchmark Raga Identification datasets and demonstrate its performance in categorizing previously seen, unseen, and all Raga classes. The proposed approach surpasses the previous NCD-based pipeline even in discovering the unseen Raga categories, offering new insights into representation learning for IAM tasks.
\end{abstract}


\section{Introduction}
Indian Art Music (IAM) is fundamentally organized around the concept of the Raga, a melodic framework that defines characteristic pitch sequences\cite{Serra_computational_2017_raga}, ornamentation styles\cite{kumarsumit2025recognizing_ornaments}, and expressive phrases\cite{phrase_based_raga}. Unlike fixed-scale systems, Ragas are defined not only by their scale but also by nuanced melodic movements, temporal variation, and improvisational trajectories~\cite{Serra_computational_2017_raga}. These features make Raga classification a challenging computational problem: the same Raga can be rendered with significant stylistic diversity, while different Ragas may share overlapping pitch material but differ subtly in ornamentation or phraseology. 
In addition to these inherent complexities, real-world and musicological settings introduce another important challenge: the presence of unseen or newly encountered Ragas. Performance archives, teaching contexts, and community-driven music repositories often contain Ragas that may not be represented in training collections or appear under sparse documentation. Thus, practical Raga identification systems must be capable not only of recognizing known Ragas but also of detecting and meaningfully organizing Ragas that the model has never encountered during training. 
This leads to a central research question: how can we design systems that reliably identify known Ragas while also discovering and structuring previously unseen Raga classes?

Unsupervised clustering and other MIR-driven discovery methods address the lack of labels but overlook the structure provided by known classes, resulting in representations that fail to leverage available supervision. 
There have been recent works utilizing Novel Class Discovery (NCD)\cite{singh2025identificationclusteringunseenragas} that attempt to bridge this gap by clustering unseen Ragas while distinguishing them from known ones. 
However, NCD-based frameworks typically train only on the unseen classes, treating them as a disjoint set from the labelled ones~\cite{hsu2019_ICLR_multiclass_without_labels_ncd,Han2019_ICCV_deep_transfer_clustering_ncd,Zhong_2021_CVPR_NCD_NCL}. 
Consequently, as the model adapts to novel categories, it often distorts previously learned representations, resulting in the phenomenon known as catastrophic forgetting\cite{joseph2022novel_ncdwf}.

To address this limitation, we adopt the Generalized Category Discovery (GCD) framework~\cite{vaze2022generalized_GCD_base}, which integrates supervised and unsupervised contrastive learning objectives within a shared embedding space, thereby jointly learning from both labeled and unlabeled data.  
In our adaptation, the encoder is trained using a combination of supervised contrastive loss on labelled data and unsupervised contrastive loss applied jointly to both labelled and unlabelled samples, while disregarding the true labels of the labelled instances. 
During unsupervised training, positive and negative pairs are constructed using structural cues such as shared audio-source identity and similarity in embedding space, so that the learned relations remain consistent with the underlying class structure.  
This design enables the model to leverage hard positives and negatives for representation learning while implicitly preserving the true class boundaries.
This joint optimization enables the model to preserve previously learned representations while simultaneously discovering novel categories—thereby mitigating catastrophic forgetting.

The main contributions of this work are as follows:
\begin{itemize}
    \item  A unified framework for identifying and organizing previously unseen Raga classes while maintaining strong performance on known Raga classes.

\item A cross-dataset evaluation on PIM and Saraga, providing insights into robustness, transferability, and domain variability.

\item An empirical improvement over our earlier NCD-based approach, supported by quantitative benchmarks and qualitative analyses of discovered Raga groups.
\end{itemize}

The codes will be shared through GitHub in the final published version.

\section{Related Works}

Early work on Raga Identification by \cite{bidkar12north_2021} employed features such as pitch and Mel-frequency Cepstral Coefficients (MFCC) on a self-curated dataset comprising 20-second audio snippets spanning 12 Raga classes.  
\cite{multimodal_HCM} utilizes multimodal information from audio and metadata to classify 2 Ragas. Another work \cite{clayton2022raga_multi_modal_Preeti_rao} explores the use of both audio and video modalities for Raga classification.  
\cite{sharma2021classification} compares multiple architectures, including Convolutional Neural Networks(CNN), Recurrent Neural Networks(RNN), and hybrid models using MFCC, spectrogram, and scalogram features, and concludes that a CNN--RNN configuration provides the most balanced performance across Hindustani and Carnatic styles.  
The work \cite{10530796} proposes a stacked ensemble framework that combines Multi Layer Perceptron(MLP), Support Vector Machine(SVM), K-Nearest Neighbour(KNN), and Random Forest classifiers, demonstrating that ensemble averaging improves the recognition of Melakartha and Janya Ragas.  
\cite{phononet_2019} utilizes a hierarchical deep learning framework called PhonoNet, which first utilizes a CNN to classify short audio segments, and subsequently applies a two-stage recurrent Long Short-Term Memory (LSTM) architecture to aggregate temporal information and classify Raga classes.  
Another work \cite{param_NCC_ontology} leverages singing times of a Raga to design a hierarchical clustering model for Raga grouping and Identification.  
The work \cite{Madhusudhan2024DeepSRGMS_raga} integrates CNN–LSTM architectures with attention mechanisms to learn hierarchical melodic patterns and produce musically meaningful similarity rankings for automatically classifying and ranking melodic sequences in Indian Classical Music.

More recently, \cite{param_XAI_TASLP_2025} uses a CNN–LSTM model trained for closed-set Raga identification over 12 Raga classes.  
We adopt the same CNN–LSTM configuration as our feature extractor.  
The work \cite{WIMAGA_confidence_Raga_ICASSP_2025} incorporates uncertainty estimation into the classification pipeline to provide reliability measures for Raga identification.  
Despite their promising performance, all these methods share a key limitation: they assume a closed-set scenario, where every test-time input belongs to one of the classes seen during training. 
Such models lack the ability to detect previously unseen Ragas, restricting their practical applicability to real-world music collections where new or rare Ragas often appear.

The work \cite{singh2025identificationclusteringunseenragas} specifically addresses this issue.  
Drawing inspiration from the NCD literature \cite{Han2019_ICCV_deep_transfer_clustering_ncd,hsu2019_ICLR_multiclass_without_labels_ncd,Zhong_2021_CVPR_NCD_NCL}, it combines contrastive, binary cross-entropy, and consistency-based losses to cluster unseen Raga classes in IAM.  
However, as in conventional NCD frameworks, this approach assumes that the unlabelled dataset is strictly disjoint from the labelled one.  
When this assumption is violated, the model suffers from catastrophic forgetting, i.e., degradation of performance on known Ragas as the model adapts to new ones\cite{joseph2022novel_ncdwf}.  

To mitigate catastrophic forgetting, several works have been proposed in the broader vision community.  
For example, \cite{joseph2022novel_ncdwf} introduced a Novel Class Discovery without Forgetting (NCDwF) framework that integrates mutual-information regularization with pseudo-latent representations to maintain consistency of known-class embeddings.  
Another work \cite{vaze2022generalized_GCD_base} presents the Generalized Category Discovery (GCD) paradigm using contrastive representation learning with Vision Transformers.  
Their approach jointly optimizes supervised and unsupervised contrastive losses across labelled and unlabelled data, followed by semi-supervised clustering to identify both known and novel classes simultaneously.

Building upon these advancements, our work adapts the GCD paradigm to the domain of IAM.  
To the best of our knowledge, this is the first study to introduce the GCD framework for Raga identification, extending prior NCD-based approaches by jointly optimizing supervised and unsupervised contrastive objectives tailored to the melodic and structural properties of Ragas.

\section{Methodology}
\begin{figure}[htbp]
\centerline{\includegraphics[width=\columnwidth]{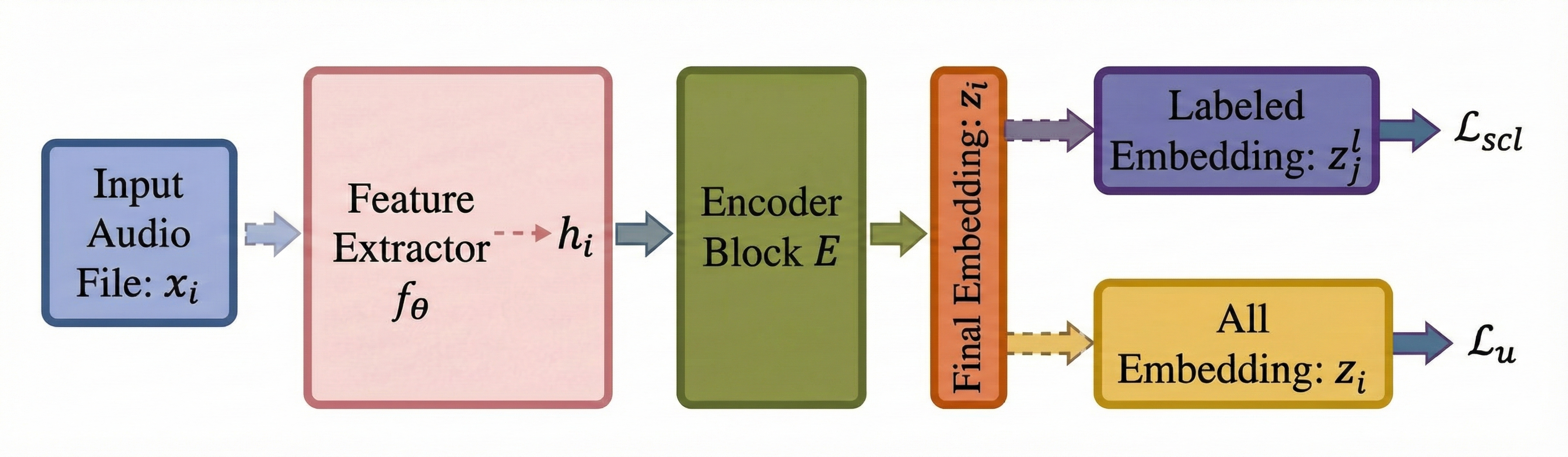}}
\caption{Overall pipeline of the proposed framework.} 
\label{fig:overall_flow}
\end{figure}

The overall flow of our methodology is shown in figure~\ref{fig:overall_flow}. The input audio clips are processed by the feature extractor \( f_{\theta} \) and encoder \( E(\cdot) \) to obtain embeddings \( z_i \). The encoder is trained on a unified objective to create meaningful clusters in the latent space.

\subsection{Dataset}
Let the labelled and unlabelled datasets be denoted as
\[
\mathcal{D}^l = \{(x_i^l, y_i^l)\}_{i=1}^{N_l}, \quad 
\mathcal{D}^u = \{x_j^u\}_{j=1}^{N_u}
\]
where \( x_i^l \) and \( x_j^u \) are audio segments from the labelled and unlabelled sets respectively, and \( y_i^l \in \{1, 2, \dots, C_l\} \) denotes their known class labels.  
Each input audio file is a 30-second clip derived from longer concert recordings.  
While true Raga labels are unknown for \( \mathcal{D}^u \), we have auxiliary metadata linking clips that originate from the same original recording.  
We represent this information by assigning each clip an identifier \( a(x) \), which denotes its source audio file.  


\begin{algorithm}[!ht]
\caption{Joint Supervised--Unsupervised Contrastive Learning for Raga Discovery}
\label{alg:gcd_raga}
\begin{algorithmic}[1]
\Require Labelled data $\mathcal{D}^l$, Unlabelled data $\mathcal{D}^u$, Feature extractor $f_\theta$, Encoder $E$, Temperature $\tau$
\Ensure Cluster assignments $\{s_i\}$ for all samples

\State \textbf{Stage 1: Feature Extraction}
\State Train $f_\theta$ on $\mathcal{D}^l$ using cross-entropy loss.
\State Obtain embeddings $h_i = f_\theta(x_i)$ for all $x_i \in \mathcal{D}^l \cup \mathcal{D}^u$.
\State Pass through encoder: $z_i = E(h_i)$.

\Statex
\State \textbf{Stage 2: Contrastive Learning}
\For{each epoch}
    \State Compute supervised contrastive loss $\mathcal{L}_{scl}$ on $\mathcal{D}^l$ 
    as shown in eq:\ref{eq:supervised}.
    \State Compute unsupervised contrastive loss $\mathcal{L}_{u}$ on $\mathcal{D}^u$ 
    as shown in eq:\ref{eq:unsupervised}.
    \State Combine losses: $\mathcal{L}_{cl} = (1-\lambda)\mathcal{L}_{scl} + \lambda\mathcal{L}_{u}$.
    \State Update encoder $E(\cdot)$ parameters via backpropagation on $\mathcal{L}_{cl}$.
\EndFor

\Statex
\State \textbf{Stage 3: Clustering and Evaluation}
\State Obtain final embeddings $\mathcal{Z} = \mathcal{Z}_l \cup \mathcal{Z}_u$.
\State Apply K-Means or cosine-similarity threshold clustering to generate cluster IDs $\{s_i\}$.
\State Evaluate performance using ACC, NMI, ARI, and Silhouette Score on Old, New, and All subsets.

\end{algorithmic}
\end{algorithm}


\subsection{Feature Extractor}

We train a CNN--LSTM classifier \( f_\theta \) on the labelled dataset \( \mathcal{D}^l \) using a cross-entropy loss over \( C_l \) classes, following the work\cite{singh2025identificationclusteringunseenragas}.  
After training converges, we remove the final softmax layer and use \( f_\theta \) as a feature extractor to obtain fixed-length embeddings:
\[
f_\theta(x_i) = h_i, \quad \text{for } x_i \in \mathcal{D}^l \cup \mathcal{D}^u
\]
The embeddings \( h_i \) are subsequently passed through a self-attention encoder \( E(\cdot) \), producing refined latent representations
$z_i=E(h_i)$.
The Encoder $E(\cdot)$ consists of a series of Encoder blocks of a transformer model with self-attention layers.
The complete embedding set is thus \(\mathcal{Z} = \{z_i\}_{i=1}^{N_l + N_u}\), which is split into \(\mathcal{Z}_l = \{z_i^l\}\) and \(\mathcal{Z}_u = \{z_j^u\}\) corresponding to the labelled and unlabelled subsets respectively. This $E(\cdot)$ is what we train during our training, while keeping the \( f_\theta \) unchanged.

\subsection{Supervised Contrastive Loss}

For each labelled embedding \( z_i^l \in \mathcal{Z}_l \), the supervised contrastive loss is defined as:
\begin{equation}\label{eq:supervised}
\mathcal{L}_{scl}^i = - \frac{1}{N_p^l} 
\sum_{z_p \in \mathcal{P}_i^l}
\log
\frac{
    \exp(\frac{\operatorname{sim}(z_i^l, z_p^l)}{\tau})
}{
   \exp(\frac{\operatorname{sim}(z_i^l, z_p^l)}{\tau})
   + 
   \sum\limits_{z_k \in \mathcal{N}_i^l}
        \exp(\frac{\operatorname{sim}(z_i^l, z_k^l)} {\tau})
}
\end{equation}
where \( \tau \) is the temperature hyperparameter controlling the sharpness of the similarity distribution.
The set of positive samples for anchor $z_i^l$ are defined as: 
\[ 
\mathcal{P}_i^l = \{z_p^l \in \mathcal{Z}_l \mid y_p^l = y_i^l, \, p \neq i\} 
\] 
These consist of all the embeddings having the same class label $y_i^l$ as that of $z_i^l$. 
\( N_p^l = |\mathcal{P}_i^l| \) represents the total number of positive samples.
To ensure efficient and meaningful optimization, we perform hard negative mining when constructing the negative.
Let the cosine similarity between any two embeddings \( z_a \) and \( z_b \) be given by:
\[
\operatorname{sim}(z_a, z_b) = \frac{z_a \cdot z_b}{\|z_a\|_2 \, \|z_b\|_2},
\]
where \( \cdot \) denotes the dot product and \( \|\cdot\|_2 \) the Euclidean norm.

For each anchor \( z_i^l \), all embeddings \( z_k^l \in \mathcal{Z}_l \) with class label \( y_k^l \neq y_i^l \) are first ranked in ascending order of cosine similarity:
\[
r(z_k^l) = \operatorname{argsort}_{k} \left( \operatorname{sim}(z_i^l, z_k^l) \right),
\]
and the set of hard negatives is defined as the 50 least similar embeddings\cite{Zhong_2021_CVPR_NCD_NCL}:
\[
\mathcal{N}_i^l = \{\, z_k^l \in \mathcal{Z}_l \mid y_k^l \neq y_i^l, \, k \in r_{1:50}(z_k^l) \,\}.
\]
This selection ensures that the loss focuses on the most challenging dissimilar examples, improving discriminative separation between Raga classes while reducing redundant gradient contributions from already well-separated negatives.

\subsection{Unsupervised Contrastive Loss}
Here, we consider the whole dataset $D^l \cup D^u$, discarding any class labels. For each embedding \( z_j \in \mathcal{Z}_l \cup \mathcal{Z}_u \), we exploit a key characteristic of our IAM datasets that each performance recording corresponds to a single Raga class. Consequently, all the derived 30-second audio clips originating from the same source audio share the same underlying Raga label. 
So, we construct positive pairs corresponding to the anchor $z_j$
constructed from clips sharing the same source identifier \( a(x_j) \) as:
\[
\mathcal{P}_j = \{\, z_p \in \mathcal{Z} \mid a(x_p) = a(x_j),\, p \neq j \,\}
\]
The set is restricted to 5 randomly sampled samples, so 
$|\mathcal{P}_j| = N_p = 5$.
The negative set \( \mathcal{N}_j \) consists of the 50 least similar samples (lowest cosine similarity) across the remaining dataset, as done for the supervised loss.
The unsupervised contrastive loss is defined as:
\begin{equation}\label{eq:unsupervised}
\mathcal{L}_{u}^j = - \frac{1}{N_p} 
\sum_{z_p \in \mathcal{P}_j}
\log
\frac{
    \exp(\frac{\operatorname{sim}(z_j, z_p)}{\tau})
}{
   \exp(\frac{\operatorname{sim}(z_j, z_p)}{\tau})
   + 
   \sum\limits_{z_k \in \mathcal{N}_j}
        \exp(\frac{\operatorname{sim}(z_j, z_k)} {\tau})
}
\end{equation}

The rationale is that, since \( f_\theta \) has been pre-trained on Raga-labelled data, the embeddings \( z_i \) are expected to preserve intra-class proximity in the latent space. Hence, clips belonging to the same Raga should remain closer in cosine similarity space, even for unlabelled data.

\subsection{Total Contrastive Objective}
The overall contrastive loss combines the supervised and unsupervised terms:
\begin{equation}\label{eq:combined_loss}
\mathcal{L}_{cl} = (1 - \lambda) 
\sum_{z_i^l \in \mathcal{Z}_l} \mathcal{L}_{scl}^i
+ 
\lambda
\sum_{z_j \in \mathcal{Z}} \mathcal{L}_{u}^j
\end{equation}
where \( \lambda \in [0,1] \) balances the contribution of labelled and unlabelled samples and is determined experimentally. 
This whole process is also explained in Algorithm~\ref{alg:gcd_raga}.
After multiple training epochs, the final embeddings from \(\mathcal{Z} = \mathcal{Z}_l \cup \mathcal{Z}_u \) are clustered to discover new Raga categories.
\begin{figure*}[ht]
\centerline{\includegraphics[width=2\columnwidth]{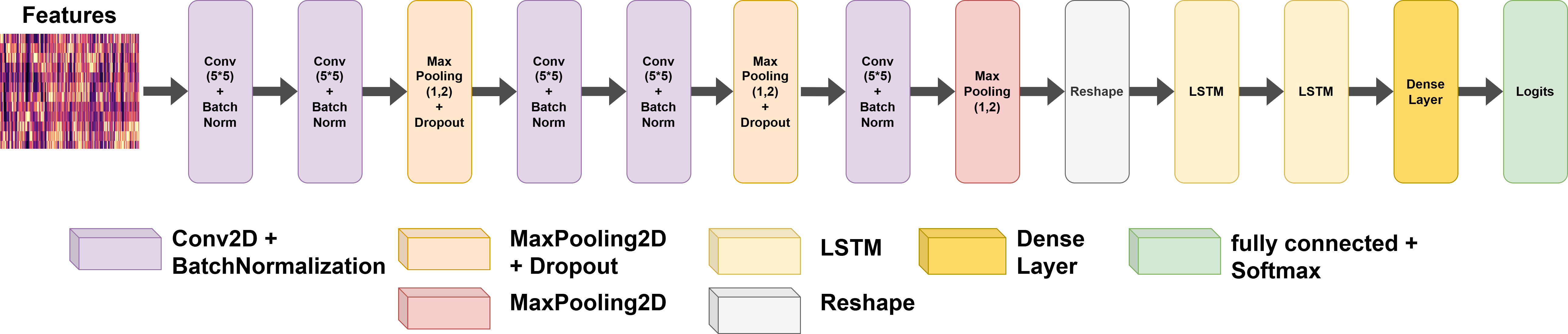}}
\caption{Architecture of the CNN-LSTM model $f_\theta$}
\label{fig:CNN_LSTM_model}
\end{figure*}
\subsection{Clustering Methods}

We employ two clustering strategies, each emphasizing a different notion of similarity in the learned embedding space.

\subsubsection*{K-means}

We apply K-Means on the learned embeddings \( z_i \in \mathcal{Z} \) to form clusters \( C_1, C_2, \ldots, C_K \). We assume that the number of classes ($k$) is known to us. Each \( z_i \) is assigned to the cluster with the nearest centroid in Euclidean distance.

\subsubsection*{Cosine Similarity}
For every pair of embeddings \( (z_i, z_j) \), we utilize cosine similarity between embeddings ($sim( (z_i, z_j)$) to directly determine cluster memberships based on pairwise proximity.
If the similarity value exceeds a predefined threshold \( \delta \), i.e.,
$\operatorname{sim}(z_i, z_j) \geq \delta$, 
The two samples are assigned to the same cluster. The clustering process continues iteratively until all samples are grouped such that intra-cluster cosine similarity is maximized and inter-cluster similarity is below the threshold.

\subsection{Evaluation Metrics}
We compute unsupervised label-dependent metrics such as clustering accuracy (ACC), Normalized Mutual Information (NMI) and Adjusted Rand Index (ARI) and a label-independent metric Silhoutte Score (SS) as explained in \cite{singh2025identificationclusteringunseenragas}, for a comprehensive evaluation of our models.


\section{Experiments}

\subsection{Datasets}

We use the Prasar Bharati Indian Music (PIM) dataset\cite{param_XAI_TASLP_2025} and the Saraga\cite{Saraga} dataset.  
The PIM dataset\cite{param_XAI_TASLP_2025} contains 191 hours of Hindustani Classical Music (HCM) recordings consisting of 501 audio recordings in 135 unique Ragas annotated with Raga and tonic labels by human experts. We select 17 Raga classes with the highest number of recordings in the dataset, where the top 12 classes form the set $D^l$ and the next 5 form the set $D^u$, same as our baseline\cite{singh2025identificationclusteringunseenragas}. $D^l$ contributes to 5702     audio clips, and $D^u$ has 2421 audio clips of 30 seconds each after segmentation.
The Saraga Hindustani\cite{Saraga} dataset is a 43-hour collection of 108 IAM recordings sung in 61 unique Ragas.
From the Saraga dataset, $D^l$ remains the same and we additionally include 5 Raga classes (1136 clips) as unlabelled data such that none of these classes overlap with those present in \( D^l \).  
The unlabelled Raga classes across PIM and Saraga may overlap with one another, but the labelled and unlabelled sets are strictly disjoint.

\subsection{Feature Extractor Selection}
We compare different feature extractors to identify the most suitable model for downstream training and clustering.
We evaluate MERT\cite{yizhi2023mert}: a transformer-based model pretrained on large-scale music data (mostly western music) using contrastive learning, CultureMERT1\cite{kanatas2025culturemert} pretrained on multicultural and multilingual music corpora, and a CNN-LSTM model~\cite{singh2025identificationclusteringunseenragas} trained from scratch on \( D^l \) using cross-entropy loss across the 12 Raga classes as shown in Figure:\ref{fig:CNN_LSTM_model}.
For each feature extractor, we obtain the corresponding embeddings for all samples in \( D^l \) and compute clustering metrics after performing K-Means clustering.  
The results show that our CNN-LSTM feature extractor consistently outperforms the pretrained transformer-based models across all metrics. So, we use our CNN-LSTM feature extractor for all subsequent experiments.

\subsection{Baseline (BL)}
As a baseline, we directly implement the NCD framework proposed by~\cite{singh2025identificationclusteringunseenragas}.  
In this approach, a feature extractor is first trained on the labelled dataset \( \mathcal{D}^l \) using its class labels, followed by training the model $E(\cdot)$ using only the $D^u$ set using a combined objective consisting of binary cross-entropy, consistency, and contrastive loss components~\cite{singh2025identificationclusteringunseenragas}.  
Since the framework is designed primarily to discover novel categories, it is expected to perform well on unseen (new) classes but may suffer degradation on previously seen classes due to catastrophic forgetting~\cite{joseph2022novel_ncdwf}.  

\subsection{Unsupervised Contrastive Learning (M1)}
In the first method (M1), the encoder \( E(\cdot) \) is trained using only the unsupervised contrastive loss term described in Eq.~\ref{eq:unsupervised}, treating the entire dataset \( \mathcal{D}^l \cup \mathcal{D}^u \) as unlabelled.  
After training, the combined embeddings \( \mathcal{Z} = \mathcal{Z}_l \cup \mathcal{Z}_u \) are clustered using K-Means to obtain predicted labels.  
This purely unsupervised setup provides insight into how effectively the encoder captures latent class structure without any supervision.  
\begin{table*}[ht]
\centering
\caption{Performance comparison of Baseline (BL), Method~1 (M1), and Method~2 (M2) across \textit{All}, \textit{New}, and \textit{Old} subsets for the PIM and Saraga datasets. The labelled set \( \mathcal{D}^l \) is always derived from PIM, while the unlabelled set \( \mathcal{D}^u \) corresponds to the respective dataset indicated. Clustering Accuracy (ACC) is reported in percentage (\%).}
\label{tab:clustering_metrics_transposed_grouped}
\scriptsize

\begin{tabular}{@{} l l 
    | c c c c 
    | c c c c 
    | c c c c @{}}

\toprule

\multirow{2}{*}{\textbf{Dataset}} &
\multirow{2}{*}{\textbf{Method}} &
\multicolumn{4}{c|}{\textbf{All}} &
\multicolumn{4}{c|}{\textbf{New}} &
\multicolumn{4}{c}{\textbf{Old}} \\
\cmidrule(lr){3-6} \cmidrule(lr){7-10} \cmidrule(lr){11-14}

& & SS & ARI & NMI & ACC & SS & ARI & NMI & ACC & SS & ARI & NMI & ACC \\

\midrule

\multirow{3}{*}{PIM}
  & BL & 0.57 & 0.51 & 0.62 & 57.96 & 0.85 & 0.64 & 0.56 & 79.34 & 0.63 & 0.71 & 0.75 & 79.24 \\
  & M1 & 0.38 & 0.52 & 0.64 & 72.62 & 0.28 & 0.54 & 0.51 & 74.27 & 0.45 & 0.76 & 0.79 & 85.01 \\
  & M2 & 0.38 & 0.65 & 0.73 & 77.49 & 0.35 & 0.71 & 0.65 & 84.68& 0.47 & 0.84 & 0.83 & 91.16  \\

\midrule

\multirow{3}{*}{Saraga}
  & BL & 0.56 & 0.50 & 0.62 & 63.42 & 0.82 & 0.44 & 0.52 & 81.04 & 0.58 & 0.54 & 0.66 & 73.64 \\
  & M1 &0.37  & 0.65 & 0.72 & 70.87 & 0.34 & 0.38 & 0.50 & 72.09& 0.42 & 0.77 & 0.79 & 85.50 \\
  & M2 &0.43  &0.69  &0.74  &74.80  & 0.30 & 0.36 & 0.48 &71.70  & 0.47 & 0.81 & 0.82 & 86.60 \\

\bottomrule

\end{tabular}
\end{table*}

\subsection{Proposed Method (M2)}
In the proposed method (M2), the self-attention encoder \( E(\cdot) \) is trained using the joint objective described in Eq.~\ref{eq:combined_loss}.  
This unified formulation enables the encoder to leverage labelled information to retain previously learned knowledge while simultaneously discovering new categories from unlabelled data.  
After training, embeddings are clustered using K-Means, as in Method~1.  
Unlike M1, which focuses solely on unsupervised representation learning, the proposed method seeks to achieve a balance between preserving performance on known classes and improving the discovery of unseen ones, thereby mitigating catastrophic forgetting.  

\subsection{Evaluation Strategy}
We evaluate all three methods using the metrics ACC, NMI, ARI, and SS.  
Performance is reported for three evaluation settings:\\  
\textit{Old} Classes: labelled subset \( D^l \) (from PIM only),\\
\textit{New} Classes: unlabelled subset \( D^u \) (from PIM and Saraga),\\
\textit{All} Classes: $D^l\cup D^u $(from PIM and Saraga)\\
allowing comparison across seen, unseen, and mixed-class scenarios.

\section{Results and Discussion}
The clustering performance for Baseline and Proposed methods over both the datasets is shown in Table:\ref{tab:clustering_metrics_transposed_grouped}
\subsection{Baseline}
For the PIM dataset, we observe that the Baseline (NCD) method performs reasonably well when classifying only the unlabelled subset (New Classes). However, its performance on the labelled subset \( D^l \) drops noticeably from 90\% accuracy to 80\%, demonstrating clear evidence of catastrophic forgetting.
The confusion matrix (not shown) highlights that the Ragas Bhairavi and Des are now predicted with less than 40\% accuracy out of $D^l$. In the \textit{All} case, the overall clustering accuracy further drops to 66.8\%. Notably, Bhopali is consistently misclassified as Shuddha Kalyan. This confusion is musically plausible, as both Ragas share similar ascending structures (\textit{Aaroh}).

For the Saraga dataset, we observe a similar trend. The baseline performs adequately on the unlabelled (New) subset but suffers sharp degradation on the labelled (Old) subset. Specifically, Bhimpilasi, Bhopali, and Marwa are among the most misclassified Ragas. Here, Bhopali belongs to \( D^l \) while Bhimpilasi and Marwa are from \( D^u \). For several labelled classes, accuracy drops below 40\%, again reflecting catastrophic forgetting. The results (shown in Table~\ref{tab:clustering_metrics_transposed_grouped}) confirm that while the baseline can effectively cluster unseen Ragas, it fails to retain learned representations for previously seen categories.

\subsection{M1}
Here, we observe a modest improvement in the performance of the \textit{Old} and \textit{All} subsets for both datasets. The encoder trained under this loss retains some latent structure from the feature extractor, resulting in slightly higher stability on known classes. However, the performance on the \textit{New} subset is either comparable to or marginally worse than the baseline NCD model. The same trends are observed across both PIM and Saraga datasets. This shows that unsupervised contrastive learning alone is insufficient to improve discovery of new categories, although it helps mitigate complete forgetting of known ones, which can be attributed to improved performance for \textit{Old} and \textit{All} cases.

\begin{figure}[htbp]
\centerline{\includegraphics[width=1\columnwidth]{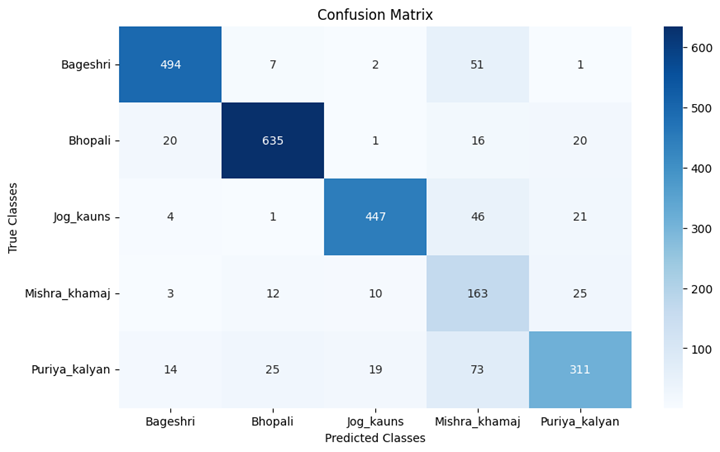}}
\caption{Confusion Matrix for M2 on $D^u$ for PIM dataset}
\label{fig:confusion matrix}
\end{figure}

\subsection{M2}
The proposed Method achieves substantial performance improvements across all subsets. On the PIM dataset, the performance on the \textit{New} subset surpasses that of the baseline NCD, while the accuracy on the \textit{Old} subset becomes even better than that achieved by the standalone feature extractor (90\%). This indicates that the joint training objective successfully preserves representations for known classes while enhancing generalization to unseen ones.
The confusion matrix in Figure~\ref{fig:confusion matrix} shows the performance on the PIM dataset for the $D^u$ subset. Clearly, it resolves a significant amount of confusion that was observed for NCD\cite{singh2025identificationclusteringunseenragas} in Raga Jog-Kouns and Puriya-Kalyan. It even improves the performance of the already well-classified Shuddha Ragas, while also enhancing performance on the Mishra Ragas (mixed Ragas).
For the \textit{All} subset, accuracy improves considerably compared to both the BL and M1, demonstrating that M2 provides a more balanced clustering across known and novel Ragas.  One thing that is consistently observed in all the Confusion matrices is a high level of confusion between Bhopali and Shuddha-Kalyan across all three settings, which is musically justifiable due to a similar melodic ascent. Another misclassification occurs between Khamaj and Mishra-Khamaj, where the latter achieves an accuracy of under 30\%, likely due to overlapping tonal material and shared note phrases.

On the Saraga dataset, M2 achieves strong performance across all evaluation settings. Here also, Bhopali and Shuddha-Kalyan exhibit partial overlap, while Marwa remains somewhat misclassified. Nonetheless, all other classes show improved cluster purity and significantly higher ACC and NMI scores, demonstrating that the model generalizes effectively across datasets with differing recording styles and acoustic conditions.

An interesting observation is that the SS for the BL is consistently higher than that of M1 and M2.  
This can be attributed to the use of a combination of contrastive, consistency, and binary cross-entropy loss (BCE) during training.  
Such a formulation likely pushes the embeddings so far in the latent space, leading to well-separated and compact clusters.  
However, this increased separation does not necessarily imply that the formed clusters align with the true underlying Raga classes.  
Consequently, while the BL model achieves higher SS values, it performs worse on label-dependent metrics ACC, NMI, and ARI, which evaluate alignment with ground-truth labels.
We also observe that clustering based on cosine similarity yields slightly lower performance but follows similar trends, whereas K-Means produces more stable and well-defined clusters for both the datasets.

\section{Conclusion and Future Scope}

In this work, we introduce a method for Raga identification and clustering, adapting the Generalized Category Discovery (GCD) paradigm.  
By combining supervised and unsupervised contrastive learning objectives, the proposed method effectively balances the retention of known-class knowledge with the discovery of previously unseen Ragas.  
Experiments conducted on benchmark datasets demonstrate that, unlike traditional Novel Class Discovery (NCD) methods, our approach mitigates catastrophic forgetting while achieving significant improvements across both known and novel class evaluations.  
The results highlight that the model preserves high accuracy for labelled classes and exhibits strong generalization in identifying new Ragas, even under different datasets.
    
Future efforts could incorporate multimodal inputs such as video or symbolic notations to enhance representation learning and disambiguate tonally similar Ragas.  
Further, integrating self-supervised or continual learning strategies may improve scalability to larger music archives while reducing reliance on labelled data.  
Expanding this approach to include Raga similarity modeling, Raga genealogy analysis, and cross-cultural melodic discovery could offer deeper insights into the latent structure of Indian Art Music and its relationship with other world music traditions.

\bibliographystyle{ieeetr}
\bibliography{referrences}

\end{document}